**Establishing a national master's programme in high-energy physics: lessons from a Romanian case study**

Roxana Zus [a], Călin Alexa [a*†e], Paul Grăvilă [b], Zsolt-Iosif Lázár [cf] and Daniel Radu [d]

[a] *University of Bucharest, Faculty of Physics, RO*
[b] *West University of Timişoara, Faculty of Physics and Mathematics, RO*
[c] *Babes-Bolyai University, Cluj-Napoca, Faculty of Physics, RO*
[d] *Alexandru-Ioan Cuza University of Iași, Faculty of Physics, RO*
[e] *IFIN-HH, Particle Physics Department, Magurele, RO*
[f] *Transylvanian Institute of Neuroscience, Cluj-Napoca, RO*

* Corresponding author: calin.alexa@cern.ch
† Associated with University of Bucharest

# Establishing a national master’s programme in high-energy physics: lessons from a Romanian case study

The establishment of a joint coordinated national programme for graduate education represents a significant step in strengthening higher education systems and their alignment with international research environments. This paper examines the creation of the Romanian National Master’s Programme in High-Energy Physics, the first initiative of its kind in the country. We outline the motivation for its development, based on the need to prepare students for active participation in large-scale international collaborations such as CERN, while also consolidating domestic expertise. The article describes the program’s design, including its multi-institutional structure, curriculum, and integration of research training, as well as the policy and administrative processes that enabled its implementation. Initial outcomes highlight the program’s contribution to student mobility, research engagement, and capacity building within the Romanian higher education system. Finally, we discuss lessons learned and future perspectives, situating the initiative within broader European efforts to modernise graduate education in the sciences.

## 1. Introduction

Graduate education plays a central role in preparing future researchers, academics, and professionals to contribute to both national development and international scientific communities. In fields such as high-energy physics (HEP), where large-scale collaborations and highly specialised expertise are required, the quality and structure of postgraduate training are especially critical (Bitter et al., 2022). In Romania, however, opportunities for specialised graduate education in high-energy physics remained limited. While Physics Master's programmes were offered by several universities, dedicated training in contemporary HEP - particularly in experimental particle physics - was available only to a very limited extent (Zus, 2026) and relied heavily on the expertise of local research groups and their participation in international collaborations.

To address these challenges and strengthen the national capacity for participation in global HEP projects, all Romanian Faculties of Physics joined forces to create the National Master's Programme in High-Energy Physics (NMP-HEP): the Faculty of Physics of the University of Bucharest (UB), the Faculty of Physics of “Alexandru Ioan Cuza” University of Iași (UAIC), the Faculty of Physics of the West University of Timișoara (UVT), and the Faculty of Physics of Babeș-Bolyai University of Cluj-Napoca (UBB) (University of Bucharest et al., 2025). The initiative was developed in close partnership with two national research and development institutes, the “Horia Hulubei” National Institute for Physics and Nuclear Engineering (IFIN-HH), Măgurele, and the National Institute for Research and Development of Isotopic and Molecular Technologies (INCDTIM), Cluj-Napoca, whose scientific expertise, research infrastructure, and human resources underpin the programme's educational and research activities (Alexa et al., 2025).

This initiative (University of Bucharest et al., 2025) represents not only a structural innovation in the national higher education system but also a strategic alignment with broader European efforts to modernise and internationalise graduate training in the sciences.

The principal contribution described in this paper is the development of a national inter-university framework model for sustaining and strengthening graduate education in high-energy physics in the context of declining student enrolment.

To our knowledge, this is the first published case study describing the establishment of a nationally coordinated Master's programme in high-energy physics involving multiple universities and national research institutes.

The remainder of this paper is organised as follows. Section 2 presents the background, motivations, and objectives that led to the establishment of the National Master’s Programme in High-Energy Physics, together with a brief comparison with similar educational initiatives in Europe and worldwide. Section 3 describes the design and structure of the program, including its curriculum, integration with international collaborations, research training opportunities, and governance framework. Section 4 discusses the implementation process, with particular emphasis on curriculum development, institutional approval procedures, governance mechanisms, and challenges encountered during implementation. Section 5 summarises the initial outcomes of the programme and examines its expected educational and scientific impact. Finally, Sections 6 and 7 present future perspectives and concluding remarks.

## 2. Background, motivation and objectives

*Background*

High-energy physics (HEP) is among the most internationalised scientific disciplines, relying on large-scale research infrastructures and multinational collaborations. Participation in experiments at CERN and other international facilities requires highly specialised expertise and a continuous supply of well-trained researchers (CERN, 2026). Universities therefore play a central role in ensuring the long-term sustainability of the scientific workforce (Bitter et al., 2022), (Malik et al., 2022).

Prior to the establishment of NMP-HEP, graduate education in high-energy physics in Romania was organised independently by individual universities, typically as part of broader Master's programmes rather than dedicated HEP programmes (Zus, 2026). Examples include *Physics of atoms, nucleus, elementary particles and astrophysics and applications, Theoretical and Computational Physics, and Astrophysics, Elementary Particles and Computational Physics*. These programmes generally enrolled a limited number of students and reflected the expertise and infrastructure available locally. As a result, curricula could not always provide comprehensive coverage of topics that have become central to contemporary collider physics, including advanced detector technologies, large-scale data analysis, distributed computing, and AI-based methods. Opportunities for laboratory training and direct exposure to international research environments also varied substantially among institutions, resulting in fragmented educational provision and limited national coordination.

At the same time, Romanian universities and research institutes have maintained a long-standing presence in major international collaborations, particularly at CERN (Neacşu, 2026). Researchers from the participating institutions contribute to several experiments and detector-development projects, creating a sustained demand for highly qualified graduates capable of working in advanced research environments. Ensuring the continuity of this participation requires educational programmes that provide students with strong theoretical foundations, exposure to modern experimental methods, computational and data-analysis skills, and direct contact with active research groups.

Figure 1 presents the evolution of personnel involved in CERN-related research activities in Romania during the period 2019–2025 (Neacşu, 2026). The data include academic staff, researchers, technical personnel, and early-career researchers participating in international collaborations associated with CERN. The categories CS I–III correspond to the Romanian national research career grades (from senior to junior researchers), AsUniv denotes university assistants, and ACS/PhD/MSc groups research assistants, doctoral students, and Master's students.

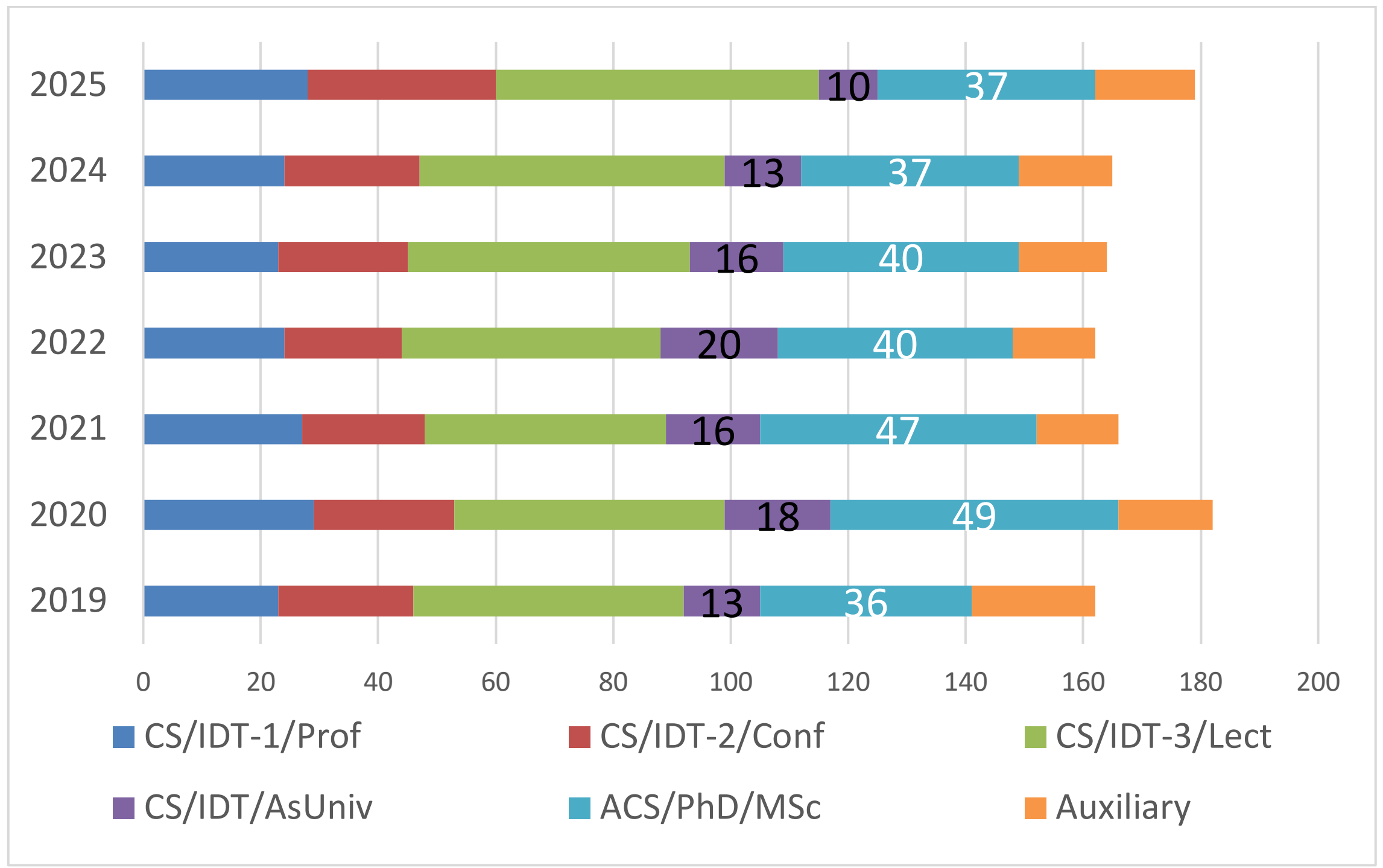

*Figure 1 Personnel participating in CERN-related research activities in Romania during the period 2019-2025, grouped by academic position and career stage.*

Of particular relevance to the present study are the ACS/PhD/MSc and AsUniv categories, which together represent the population of young researchers and students actively engaged in research and at the beginning of their academic careers. The sustained participation of these early-career researchers demonstrates an active educational and research pipeline supporting the long-term development of high-energy physics in Romania. This human-resource base provides a strong foundation for the National Master's Programme in High-Energy Physics and highlights the importance of coordinated educational initiatives aimed at preparing future generations of researchers for participation in international scientific collaborations.

A further challenge emerged from the gradual decline in student enrolment in physics programmes observed over recent years. The decreasing number of students interested in highly specialised fields such as particle physics raised concerns regarding the long-term sustainability of local graduate programmes. Individually, the participating universities faced difficulties in maintaining a sufficiently large and diverse student cohort while offering a comprehensive curriculum covering both theoretical and experimental aspects of the discipline.

Figure 2 presents the evolution of student enrolment in Physics programs at the four universities participating in the National Master's Programme in High-Energy Physics (NMP-HEP) during the period 2016–2025. The upper panels show the numbers of Bachelor's and Master's students, while the lower panels display the numbers of PhD students and the corresponding national totals aggregated across all participating institutions. The data reveal substantial differences in the size and dynamics of Physics programs among the partner universities, reflecting variations in institutional profiles, local recruitment capacity, and available educational resources.

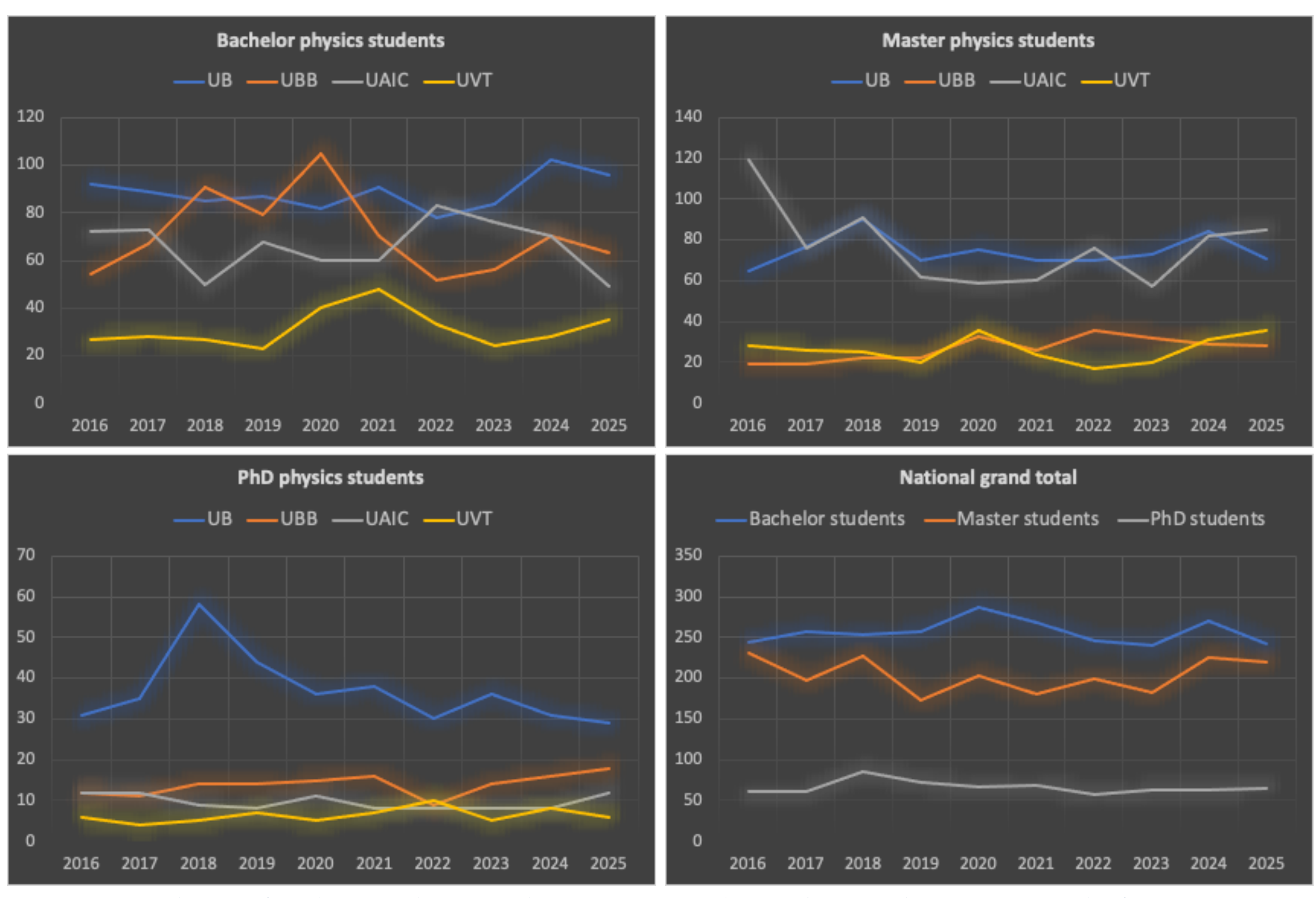

*Figure 2 Evolution of student enrolment in physics programs during the period 2016–2025 at the four universities participating in the National Master's Programme in High-Energy Physics (NMP-HEP). The upper panels show the number of Bachelor's (left) and Master's (right) students, while the lower panels present the number of PhD students (left) and the national totals aggregated over all participating institutions (right).*

At the national level, undergraduate enrolment remains relatively stable throughout the period considered, indicating a sustained interest in Physics studies. In contrast, the number of students pursuing advanced graduate education is considerably smaller, particularly at the doctoral level. Although Master's enrolment remains relatively stable at the national scale, the limited size of graduate student populations highlights the challenges associated with maintaining a continuous educational pathway from undergraduate studies to research careers in physics.

An additional challenge affecting graduate education in physics in Romania arises from institutional regulations concerning the minimum number of students required for a Master's programme to operate. In practice, programmes taught in Romanian typically require a minimum enrolment of approximately 11-12 students, while programmes taught in English generally require at least 8 students. Furthermore, these minimum enrolment requirements are scheduled to increase from the next academic year in order to improve the financial viability of Master's programmes, making it even more difficult for highly specialised programmes with relatively small student populations to operate independently. At the same time, the number of publicly funded places allocated to individual Master's programmes is often lower than these thresholds. Given the relatively small number of applicants to highly specialised fields, these conditions may prevent certain programmes from being offered in a given academic year, even when student demand exists. As a result, students may be required to enrol in alternative Master's programmes that have achieved the required enrolment, even when these programmes do not correspond to their original academic interests or career aspirations.

For highly specialised disciplines such as high-energy physics, these conditions may disrupt the educational pathway from undergraduate studies to doctoral research and compromise the long-term sustainability of the field. The establishment of a national inter-university Master's programme therefore represents a mechanism for consolidating student demand and ensuring the continuity of advanced training in strategically important scientific disciplines.

*Motivation*

The creation of the National Master’s Programme in High-Energy Physics (NMP-HEP) (University of Bucharest et al., 2025) was motivated by two complementary considerations. The first was the active and long-standing participation of all four partner universities in CERN experiments and other international high-energy physics collaborations. Maintaining and expanding this involvement requires a continuous supply of highly qualified graduates equipped with the theoretical knowledge, experimental skills, and research experience necessary to contribute effectively to large-scale scientific projects. The participating institutions therefore recognised the importance of strengthening graduate education in order to enhance both the visibility and the long-term sustainability of Romanian contributions to international research.

A second and equally important motivation was the significant decline in student enrolment in physics programmes observed across the participating universities. The decreasing number of students pursuing advanced studies in highly specialised fields such as particle physics posed a challenge to the sustainability of local Master’s programmes and limited the ability of individual institutions to offer a comprehensive and diverse curriculum. Under these circumstances, maintaining separate programmes at each university was becoming increasingly difficult from both academic and organisational perspectives.

In response to these challenges, the partner institutions adopted a collaborative approach based on the development of a unified national Master's programme. By pooling academic expertise, educational resources, and research infrastructure, the NMP-HEP programme provides students with access to a curriculum that reflects the educational requirements of contemporary high-energy physics (University of Bucharest et al., 2025). The programme combines advanced theoretical and experimental training with modern computational and data-analysis methods, while also offering research opportunities and mobility experiences across participating institutions. A distinctive feature of the programme is the active involvement of two national research and development institutes, IFIN-HH and NCDTIM, which make available their research infrastructure, scientific expertise, and human resources in support of teaching and research activities. Through this integrated framework, the programme contributes to strengthening Romania's capacity to train future researchers and to sustain long-term participation in international high-energy physics collaborations.

*Objectives of the joint master's initiative*

The establishment of the joint Master's programme in High-Energy Physics (HEP) was guided by a set of common objectives shared by the partner institutions. The primary objective was to create a sustainable national framework for graduate education in high-energy physics, capable of addressing declining student enrolment while ensuring the continued availability of advanced training in the field. A second objective was to develop a coherent curriculum that reflects the competencies and skills required for participation in contemporary high-energy physics research.

An equally important objective was to strengthen cooperation among the participating universities and research institutes through the adoption of common academic strategies, coordinated teaching activities, and mechanisms for the continuous monitoring and evaluation of the programme. The partnership was also intended to facilitate the sharing of expertise, research infrastructure, and educational resources, thereby providing students with learning opportunities that would not be available within isolated institutional frameworks.

The programme further sought to promote national and international collaboration in high-energy physics, particularly through closer integration with major research infrastructures and international scientific collaborations. Supporting and stimulating students interested in careers in high-energy physics constituted another key objective, pursued through specialised training, research projects, internships, and mobility opportunities across the participating universities, as well as through access to the research infrastructure and scientific expertise available at the partner national research and development institutes. Finally, the programme aimed to promote the exchange of good educational practices among academic staff and to contribute to the continuous improvement of teaching and learning standards within the participating institutions.

*Benchmarking with similar programs in Europe and worldwide*

The development of the National Master's Programme in High-Energy Physics (NMP-HEP) was informed by existing models of graduate education in particle physics offered by leading universities and international consortia. Across Europe, several institutions provide specialised Master's programmes that combine advanced theoretical and experimental training with direct exposure to major international research infrastructures and collaborations.

One of the closest international examples is the Erasmus Mundus Joint Master in Advanced Methods in Particle Physics (IMAPP) (University of Clermont Auvergne et al., n.d.), a two-year, 120 ECTS programme jointly delivered by TU Dortmund University (Germany), the University of Bologna (Italy), and Université Clermont Auvergne (France). The programme emphasises student mobility, inter-institutional cooperation, and advanced training in particle physics. Similar examples include the Master's programme in Experimental Particle and Nuclear Physics at Lund University (Sweden) (Lund University, n.d.), the specialization in Particle Physics and Astrophysical Sciences at the University of Helsinki (Finland) (University of Helsinki, n.d.), and the Master's programme in Astrophysics, Particle Physics and Cosmology at the Universitat de Barcelona (Spain) (University of Barcelona, n.d.). These programmes illustrate the growing importance of international collaboration, research-based learning, and mobility in graduate education in fundamental physics.

Outside Europe, high-energy physics education is typically organised through graduate programmes offered by individual research-intensive universities. Examples include programmes at the University of Chicago, the University of Michigan, the University of Tokyo, and Tsinghua University, all of which maintain strong links to major international particle-physics collaborations. While these programmes provide excellent training opportunities, they are generally developed and administered by a single institution rather than through formal national partnerships.

Against this international landscape, the Romanian National Master's Programme in High-Energy Physics presents a distinctive organisational model. The programme is jointly delivered by four universities - the University of Bucharest, Alexandru Ioan Cuza University of Iași, West University of Timișoara, and Babeș-Bolyai University of Cluj-Napoca - with the active participation of two national research and development institutes, the Horia Hulubei National Institute for Physics and Nuclear Engineering (IFIN-HH) and the National Institute for Research and Development of Isotopic and Molecular Technologies (INCDTIM) (University of Bucharest et al., 2025). The partnership is based on a common curriculum, shared governance, coordinated mobility mechanisms, and access to research infrastructure distributed across multiple institutions.

Unlike these examples, NMP-HEP was conceived not primarily as an international mobility programme, but as a national strategy for preserving and strengthening graduate education in a highly specialised discipline through coordinated inter-institutional cooperation (Alexa et al., 2025).

## 3. Design and structure of the Master's programme

The National Master's Programme in High-Energy Physics (NMP-HEP) is an English-taught graduate programme established within a national framework of academic cooperation. The programme awards the qualification *High-Energy Physics (HEP)* and is structured over two academic years, comprising a total of 120 ECTS credits in accordance with the European Higher Education Area (University of Bucharest et al., 2025). Its structure is designed to combine advanced academic training with access to the complementary expertise, educational resources, and research environments available across the partner institutions.

*Institutional coordination and governance*

Admission to the programme is organised independently by each partner university, where candidates apply and complete the selection process. Students are enrolled at the institution to which they are admitted, while the overall student body of the National Master's Programme in High-Energy Physics is formed through the aggregation of admissions across all participating institutions. To support the effective implementation of the program, each partner appoints a local academic coordinator responsible for facilitating communication, overseeing academic activities, and ensuring cooperation within the consortium. The coordinators meet regularly to coordinate student mobility, monitor the implementation of the curriculum, discuss proposed curricular developments, and address issues related to the operation of the programme. Through this governance structure, the participating institutions maintain a coherent academic framework while preserving their institutional autonomy.

*Educational activities and student support*

The partner universities provide all students enrolled in the HEP Master's programme with access to the educational infrastructure required for their studies. In addition, the partner national research and development institutes contribute through access to research facilities, scientific expertise, and opportunities for participation in ongoing research activities. Teaching activities are delivered by tenured and affiliated academic staff from the participating national R&D institutes, ensuring broad coverage of the curriculum and exposure to diverse areas of expertise.

Student mobility constitutes an important component of the programme and is coordinated by the universities of enrolment in collaboration with the host institutions. To facilitate participation in mobility activities, host universities provide accommodation in their student residences. These arrangements enable students to benefit from the complementary educational and research resources available across the partnership while maintaining continuity in their academic progression. Collectively, these measures create an integrated learning environment that supports both student development and inter-institutional cooperation.

*Rights and responsibilities of partner institutions*

The partner institutions share responsibility for ensuring appropriate academic and administrative support for students throughout their studies and periods of mobility. Host institutions provide the necessary conditions for academic integration and participation in local activities, while home institutions are responsible for preparing and guiding students prior to mobility periods. The partnership is also based on a shared commitment to the continuous development of the programme. Accordingly, each institution may propose modifications to the curriculum or organisational framework, with any changes subject to consultation and collective approval by all partners and implemented in accordance with institutional and national regulations, as *Romanian Higher Education Law No. 199/2023* (2023) and established by Romanian Agency for Quality Assurance in Higher Education (ARACIS) (2025). These arrangements establish a framework of shared responsibility, transparency, and mutual accountability that supports the long-term operation of the programme.

*Evaluations and completion of studies*

Student assessment is conducted in English, the language of instruction of the program, and is organised by the institution responsible for delivering the corresponding course or educational activity. To ensure the effective operation of the common curriculum, all partner institutions recognise and accept the assessment results awarded by the other members of the consortium. This mutual recognition framework facilitates student mobility while maintaining consistent academic standards across the programme.

The completion of studies includes the preparation and defence of a dissertation carried out under the academic regulations of the institution of enrolment. Dissertation committees are established in consultation among the partner institutions and may include academic staff from multiple universities, reflecting the collaborative nature of the programme. Although graduation procedures are administered by the institution where the student is enrolled, the participation of faculty members from across the consortium contributes to the academic coherence and quality assurance of the degree.

*Final provisions*

The operation of the NMP-HEP programme is governed by a formal partnership agreement that defines the responsibilities of participating institutions, procedures for programme development, and mechanisms for institutional cooperation. The agreement includes provisions for periodic review and amendment, the admission of new partners, and procedures governing institutional withdrawal. To ensure the stability of the programme and protect students' academic progression, any modifications to the partnership framework are implemented in a manner that preserves the rights and obligations of students already enrolled. Disputes arising within the partnership are addressed through consultation and mutual agreement among the participating institutions. These provisions provide a stable and flexible framework for the long-term sustainability of the programme.

*Curriculum overview*

The curriculum of the National Master's Programme in High-Energy Physics (NMP-HEP) (University of Bucharest et al., n.d.) was designed to provide students with a balanced combination of theoretical knowledge, experimental competencies, computational skills, and research experience aligned with the requirements of contemporary particle-physics research. Structured over four semesters and comprising 120 ECTS credits, the curriculum integrates courses offered by the participating universities with research and training activities conducted in collaboration with the partner research institutes.

The first component of the curriculum focuses on the theoretical foundations of particle physics, including relativistic quantum mechanics, quantum electrodynamics, the Standard Model, and extensions of the Standard Model. These courses provide students with the conceptual framework necessary for understanding the fundamental constituents of matter and their interactions.

A second component addresses experimental methods and detector technologies through courses dedicated to particle detectors, data acquisition systems, and experimental techniques employed in modern collider experiments. These subjects are complemented by courses that expose students to current research topics and scientific developments in high-energy physics.

The curriculum also places a strong emphasis on computational methods and data analysis. Students receive training in programming, Monte Carlo simulations, computational approaches to high-energy physics, and advanced data-analysis techniques. These topics reflect the growing importance of large-scale computing infrastructures, statistical methods, and artificial intelligence tools in contemporary particle-physics research.

Research training is integrated throughout the programme through research practice, scientific research activities, internships, and the preparation of a Master's dissertation. This progressive involvement in research enables students to apply the knowledge acquired in coursework to real scientific problems while benefiting from the expertise and infrastructure available across the participating institutions.

Figure 3 illustrates the progressive integration of theoretical and experimental training, advanced computing methods, research activities, and international scientific collaborations throughout the four semesters of the programme. The curriculum is embedded within an active research environment involving CERN experiments, phenomenological studies related to present and future colliders, and detector research and development activities, providing a pathway toward doctoral studies and research careers.

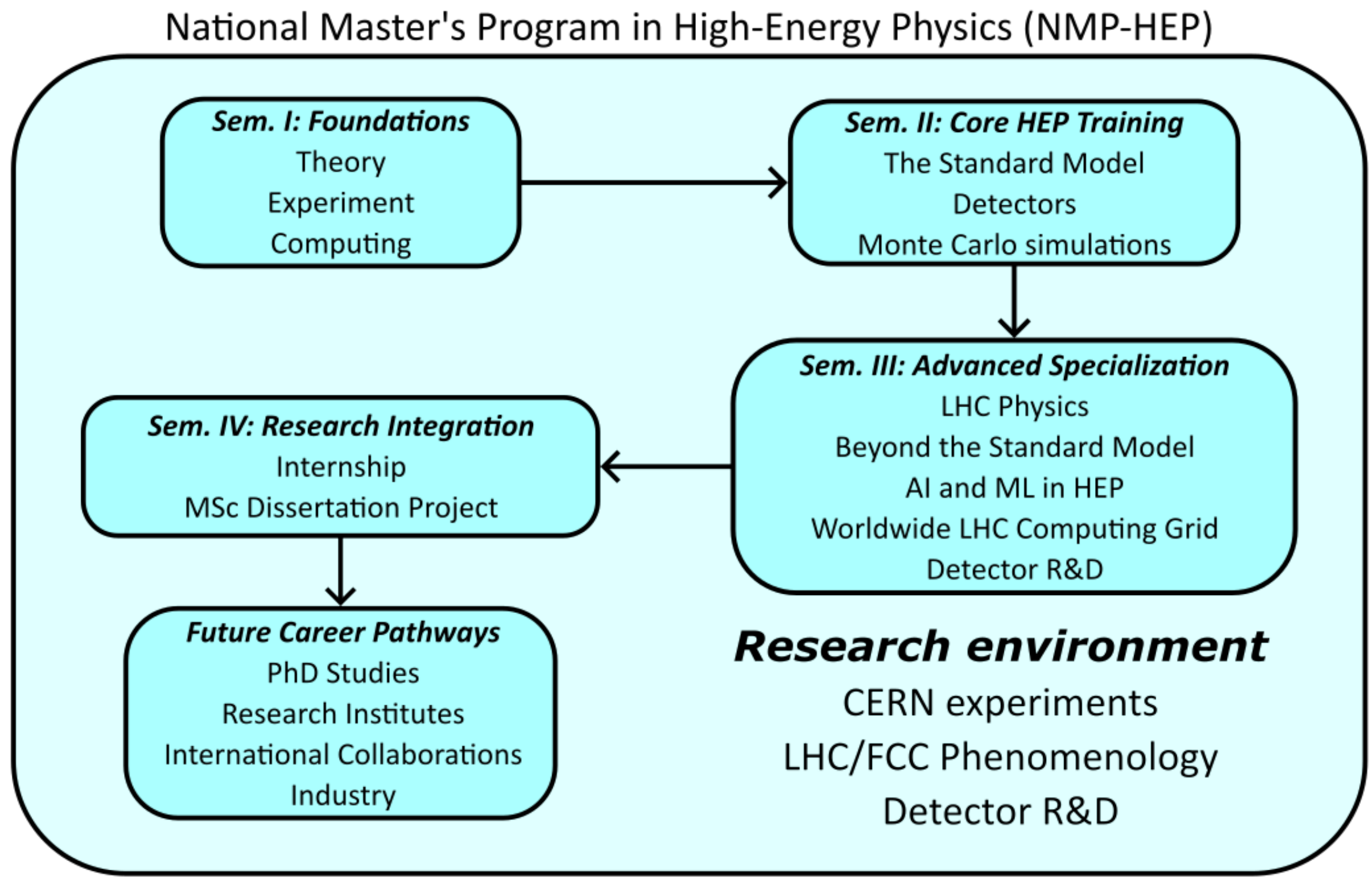

*Figure 3 Educational pathway and integration of education with research within the National Master's Programme in High-Energy Physics.*

Taken together, these curricular components provide students with a comprehensive preparation for doctoral studies and for participation in national and international research collaborations in high-energy physics.

Table 1 summarises the principal components of the NMP-HEP curriculum together with representative courses. Rather than providing an exhaustive list of modules, it illustrates the balance between theoretical foundations, experimental methods, computational training, research activities, and transferable skills that characterises the educational philosophy of the programme.

| **Curriculum Component** | **Representative Courses** |
|---|---|
| Theoretical Foundations | Relativistic Quantum Mechanics, Standard Model, BSM |
| Experimental Methods | Particle Detectors I & II, DAQ I & II |
| Computing and Data Analysis | Monte Carlo Simulations, Computational HEP, Data Analysis |
| Research Training | Research Practice, Scientific Research, Dissertation |
| Transferable Skills | Ethics in Research, Programming, VHDL |

*Table 1 Educational components and representative courses of the NMP-HEP curriculum.*

*Integration with international collaborations*

A central feature of the NMP-HEP programme is its strong integration with the international high-energy physics community. The participating universities and partner research institutes maintain active involvement in major international collaborations, particularly at CERN (Neacșu, 2026), providing students with opportunities to engage with contemporary research activities and large-scale scientific infrastructures. Through research projects, internships, mobility periods, and participation in international schools, students are exposed to the collaborative and multidisciplinary environment that characterises modern particle physics.

The programme also benefits from established partnerships with research institutions and universities abroad, creating opportunities for academic exchanges and advanced training. These include pathways for participation in specialised Master's-level studies and research placements at leading European laboratories and research centres. Such experiences complement the formal curriculum and contribute to the development of scientific, technical, and intercultural competencies that are increasingly important in international research careers.

Examples include participation in CERN-based research activities, attendance at international schools in particle physics and detector technologies, and mobility opportunities such as the Master programme hosted by CPPM Marseille. These activities provide students with direct exposure to international research environments and facilitate their integration into European scientific networks.

*Research supervision, mobility, and thesis opportunities*

Research training constitutes a central component of the NMP-HEP programme. Throughout their studies, students are encouraged to engage in research activities under the supervision of academic staff from the participating universities and researchers from the partner national research institutes. This approach enables students to benefit from a broad range of expertise and provides access to research topics aligned with ongoing national and international scientific projects.

Student mobility represents an important mechanism for expanding educational and research opportunities. Through mobility periods at partner institutions, students can attend specialised courses, participate in research activities, and gain access to complementary facilities and expertise that may not be available at their home institution. These experiences contribute to the development of both scientific competencies and collaborative skills that are essential in contemporary research environments.

The Master’s dissertation provides students with an opportunity to undertake an independent research project under academic supervision. Thesis topics are typically linked to ongoing research activities within the participating universities, partner research institutes, or international collaborations, including projects associated with CERN experiments. Through this process, students acquire experience in scientific methodology, data analysis, and research communication, while contributing to active research programs in high-energy physics.

Where appropriate, dissertation projects may be jointly supervised by academic staff from the participating universities and researchers from the partner research institutes, further strengthening the integration of education and research within the programme.

## 4. Implementation process

*Formation of the Consortium*

The development of NMP-HEP was preceded by a period of consultation among the participating universities and research institutes. These discussions identified common educational challenges, including declining enrolment, fragmented educational provision, and the need for stronger alignment with international research activities. The resulting partnership brought together complementary expertise and resources with the aim of creating a sustainable national framework for graduate education in high-energy physics.

*Development of the common curriculum*

The development of a common curriculum represented one of the central components of the NMP-HEP project. The process began with a comparative analysis of the graduate-level courses and educational practices available at the participating universities. Taking into account the different profiles, expertise, and resources of the partner institutions, the University of Bucharest, benefiting from the close involvement of researchers from the Particle Physics Department of IFIN-HH, played a leading role in shaping the core HEP curriculum, while the other partner universities contributed to the broader academic and institutional framework of the programme (Alexa et al., 2025). The objective was not to replace existing strengths, but rather to integrate them within a coherent educational framework capable of supporting advanced training in contemporary high-energy physics (Bitter et al., 2022), (Malik et al, 2022).

The curriculum was designed through a collaborative process involving academic staff from the participating universities and researchers from the partner national research institutes. Particular attention was given to aligning the programme with the educational and research requirements of contemporary high-energy physics, while ensuring compatibility with the European Higher Education Area (n.d.). The resulting curriculum combines theoretical and experimental components, computational and data-analysis training, and research-oriented activities within a unified structure of 120 ECTS credits.

An important consideration during the curriculum design process was the integration of research training. The involvement of IFIN-HH and INCDTIM enabled the inclusion of research projects, internships, and dissertation opportunities connected to active scientific programmes and international collaborations. This approach strengthened the link between formal coursework and research practice, providing students with opportunities to apply acquired knowledge in realistic scientific environments. Furthermore, students can benefit from research internships at other research institutes and centres with which the faculties have collaboration agreements, especially in Bucharest-Măgurele, as shown by the mapping of groups involved in HEP experiments (Neacșu, 2026).

When developing the curriculum, a key consideration was balancing online and face-to-face activities. Although national legislation permits up to 60% of total course hours (learning and teaching) and 35% of practical and research activities to be conducted online (ARACIS, 2022), the design also had to accommodate student mobility periods across the four partner universities. Consequently, the curriculum prioritises delivering suitable lectures and tutorials online while maximizing face-to-face engagement at the partner campuses. Where necessary, particularly for practical components, a modular delivery model was adopted. Ultimately, only two courses required this modular format, and only for a select set of lab sessions held 2–3 times per week. Where feasible, short-term teaching staff mobility offers an additional solution.

The development of the curriculum required continuous consultation among the partner institutions in order to harmonise learning outcomes, assessment procedures, and academic requirements. The resulting framework reflects a balance between institutional diversity and academic coherence, while preserving the flexibility necessary to accommodate future developments in the field.

*Institutional approval process*

The establishment of NMP-HEP required the coordination of academic and administrative procedures across all participating institutions. Following the definition of the common curriculum and governance framework, the proposed programme underwent a series of internal evaluations and approval stages within each partner university. These procedures included consultations at the departmental and faculty levels, followed by review and approval by the relevant academic bodies in accordance with institutional regulations and national higher-education requirements.

A key challenge during this process was ensuring consistency among the academic and administrative procedures of the participating institutions while preserving their institutional autonomy. Regular consultations among the partners facilitated the harmonization of programme documentation, curriculum structure, admission procedures, and mobility arrangements. The shared commitment of the participating universities and research institutes proved essential for maintaining coherence throughout the approval process.

Unlike many collaborative educational initiatives that build upon existing degree programmes, NMP-HEP was initially established as a new Master's programme and therefore required formal approval and accreditation procedures at each participating university and further, at national level. Finally, more than a year after the collaboration agreement was established, a change in national regulations allowed for recognition within the existing accredited domain. The successful completion of these processes enabled the creation of a common educational framework supported by multiple universities and national research institutes, laying the foundation for the long-term governance, operation, and sustainability of the programme. The master programme was officially included for all universities in the Government Decision No. 428/2025 (2025).

*Establishment of governance and mobility mechanisms*

Following the approval and accreditation of the program, a governance structure was established to support its implementation and long-term operation. Each participating university appointed a local academic coordinator responsible for facilitating communication among partners, overseeing the local implementation of the curriculum, and supporting student participation in programme activities. Regular meetings among the coordinators provided a mechanism for discussing academic matters, monitoring programme development, coordinating mobility activities, and addressing operational challenges.

Particular attention was devoted to the establishment of mobility mechanisms capable of supporting a distributed educational model. Since students remain enrolled at their home institution while benefiting from educational and research opportunities across the partnership, procedures were developed to facilitate the recognition of courses, assessments, and academic activities completed at partner institutions. These arrangements ensured that mobility could be integrated into students' academic pathways without creating additional administrative barriers.

The governance framework was further strengthened through the active participation of the partner national research institutes, which contributed scientific expertise, research opportunities, and access to specialised infrastructure. Together, these governance and mobility mechanisms created the institutional conditions necessary for the effective operation of a national Master's programme distributed across multiple universities and research organisations.

*Challenges encountered during implementation*

The implementation of NMP-HEP was accompanied by several challenges associated with the establishment of a novel educational framework involving multiple universities. One of the most significant difficulties arose from the limited availability of specific regulatory provisions addressing the creation and operation of national inter-university Master's programmes. While the national higher-education framework includes regulations governing conventional degree programmes and international joint programmes, fewer guidelines exist for nationally coordinated programmes involving multiple higher-education institutions according to the *Romanian Higher Education Law No. 199/2023* (2023) and the Guidelines and Specific Standards established by the Romanian Agency for Quality Assurance in Higher Education (ARACIS).

As a result, the participating institutions were required to identify practical solutions for issues related to governance, curriculum coordination, student mobility, academic recognition, and programme administration. This process involved extensive consultation among academic and administrative structures within the ministry agencies and the partner institutions, as well as careful interpretation of the existing regulatory framework to ensure compliance with national requirements.

Additional challenges were similar to the one faced by joint master programmes at European level (e.g. Leung, 2025) and included the harmonization of institutional procedures, the coordination of academic calendars, and the development of common mechanisms for curriculum management and student academic, administrative and financial support. The successful implementation of the programme therefore depended not only on scientific and educational considerations, but also on sustained institutional cooperation and a shared commitment to overcoming organisational and regulatory barriers and limitations.

Despite these challenges, the implementation process demonstrated that national inter-university cooperation can provide an effective framework for the development of graduate education in highly specialised scientific fields. The experience gained through NMP-HEP may therefore offer useful insights for similar initiatives in other disciplines facing comparable educational and demographic challenges.

*Lessons learned*

Several lessons emerged from the implementation of the National Master's Programme in High-Energy Physics. First, the existence of a shared scientific vision among the participating institutions proved essential for building consensus and maintaining long-term commitment to the project. The development of a common curriculum and governance framework required sustained dialogue and a willingness to balance institutional interests with collective objectives.

Second, the active involvement of national research institutes significantly enhanced the educational value of the programme. Beyond contributing teaching expertise, the institutes provided access to research infrastructure, ongoing scientific projects, and international collaborations, thereby strengthening the integration of education and research. This experience suggests that partnerships between universities and national laboratories can play an important role in supporting graduate education in highly specialised disciplines.

Third, the implementation process demonstrated the importance of flexible institutional cooperation. The creation of a national inter-university programme required the harmonization of academic procedures, mobility arrangements, and administrative practices across multiple institutions. Although existing regulations did not always provide explicit guidance for such a framework, sustained collaboration among the partners enabled practical solutions to be identified and implemented.

Finally, the experience of NMP-HEP indicates that national cooperation can provide an effective response to challenges associated with declining enrolment in highly specialised scientific fields. By pooling expertise, educational resources, and research opportunities, participating institutions were able to offer a richer and more sustainable educational environment than would have been possible through isolated local programs.

These lessons may be relevant to other disciplines and countries seeking to preserve advanced training capacity in strategically important but demographically vulnerable fields.

## 5. Initial outcomes

Although the National Master's Programme in High-Energy Physics (NMP-HEP) is still in the early stages of implementation, several important outcomes have already emerged.

Beyond the establishment of a new academic program, NMP-HEP has created a sustainable framework for cooperation among universities and research institutes, enabling the pooling of expertise, educational resources, and research infrastructure within a common educational model. The successful completion of the first academic year demonstrates the feasibility of this collaborative approach and provides a foundation for both immediate educational benefits and longer-term contributions to high-energy physics education and research in Romania.

At the time of writing, the National Master's Programme in High-Energy Physics has successfully completed its first academic year of operation. The common curriculum has been implemented across the participating institutions, and the inaugural class of students has completed the teaching activities associated with the first two semesters of study. The programme has therefore progressed from the planning and accreditation stages to the effective delivery of graduate education within the collaborative framework established by the partner institutions.

The first year of implementation has also enabled the validation of the governance and coordination mechanisms developed during the design phase of the programme. Regular interaction among academic coordinators, collaboration among teaching staff, and the involvement of partner research institutes have contributed to the effective organisation of educational activities. These developments demonstrate the feasibility of a national inter-university model for graduate education in a highly specialised scientific field.

Although a comprehensive assessment of the long-term impact of NMP-HEP will require several years of operation and the graduation of multiple generations of students, the successful implementation of the first year provides encouraging evidence regarding the viability of this collaborative educational model. Beyond the completion of the taught component of the curriculum, students have already begun to engage in research activities spanning a broad range of topics in high-energy physics. An initial research progress session, organised at the Faculty of Physics of the University of Bucharest on 22 May 2026, enabled students to present the status of their research projects and receive feedback from academic staff and researchers representing the participating universities and research institutes (University of Bucharest, 2026).

The first year of operation has also demonstrated the ability of the partner national research and development institutes to contribute actively to research training within the programme. Through the involvement of researchers from IFIN-HH and INCDTIM, students have gained access to research topics closely connected to ongoing scientific programmes, including participation in international experimental collaborations, phenomenological studies related to physics at the Large Hadron Collider (LHC), preparatory research activities for future facilities such as the Future Circular Collider (FCC), and detector research and development (R&D) collaborations. These early developments suggest that the programme is already fulfilling one of its central objectives, namely the integration of graduate education with active research environments at both the national and international levels.

Looking ahead, the programme is expected to strengthen the national capacity for training future researchers, enhance student participation in international scientific collaborations, and contribute to the long-term sustainability of high-energy physics education in Romania. The experience gained during the implementation of NMP-HEP may also provide a useful model for other highly specialised disciplines facing similar educational and demographic challenges.

## 6. Future perspectives

The establishment of a national framework for graduate education in high-energy physics has created new opportunities for academic cooperation, research training, and international engagement. The long-term success of the programme will depend on its ability to adapt to evolving educational and scientific needs while addressing challenges related to sustainability, funding, and student recruitment. The following perspectives outline potential directions for the continued development of the programme.

*Consolidation and expansion of the program*

The successful completion of the first academic year provides a strong foundation for the continued development of NMP-HEP. Future efforts will focus on further strengthening the curriculum through the introduction of emerging topics in particle physics, detector technologies, advanced computing, and artificial intelligence applications. Additional priorities include expanding student mobility opportunities, strengthening collaboration with international partners, and increasing participation in international schools, research internships, and exchange programs. The development of dedicated fellowship and scholarship schemes may further enhance the attractiveness of the programme and facilitate student participation in research activities.

*Contribution to research capacity and workforce development*

Beyond its educational objectives, NMP-HEP is expected to contribute to the development of highly qualified human resources capable of supporting research and innovation activities in Romania and abroad. By providing advanced training in theoretical, experimental, and computational aspects of high-energy physics, the programme prepares graduates for doctoral studies and research careers within universities, research institutes, and international organisations. In particular, the programme is expected to strengthen the pipeline of young researchers contributing to major scientific collaborations, including those hosted at CERN and other large-scale research infrastructures.

An important next step in the maturation of the programme will be the development of stronger links with doctoral education. As the first generation of NMP-HEP students approaches graduation, the participating institutions intend to expand opportunities for continuation into PhD studies, particularly within research groups involved in major international collaborations. In preparation for this transition, efforts are already underway to identify high-quality doctoral research topics associated with ongoing experimental and theoretical programmes, including activities at CERN and other leading international research infrastructures. Such a strategy is expected not only to support the academic progression of graduates but also to enhance the international visibility of the participating institutions through the active involvement of doctoral students in internationally recognised scientific collaborations.

*Long-term challenges and opportunities*

The long-term success of NMP-HEP will depend on its ability to maintain institutional commitment, secure sustainable funding, and attract talented students in a period characterised by declining enrolment in the physical sciences. Continued investment in student support, mobility opportunities, and research training will be essential for preserving the competitiveness and attractiveness of the programme. Looking ahead, the experience gained through NMP-HEP may create opportunities for deeper integration within the European Higher Education Area, including the development of joint degrees, expanded international partnerships, and potentially the establishment of a coordinated doctoral framework in high-energy physics. More broadly, the programme may serve as a model for national cooperation in other highly specialised disciplines facing similar educational and demographic challenges.

## 7. Conclusion

This paper presented the establishment of the National Master's Programme in High-Energy Physics (NMP-HEP), a collaborative educational framework developed jointly by all Romanian Faculties of Physics in partnership with two national research and development institutes. The programme was conceived in response to the challenges posed by declining student enrolment, fragmented educational provision, and the need to strengthen the preparation of future researchers for participation in international scientific collaborations. Through the development of a common curriculum, shared governance mechanisms, coordinated mobility arrangements, and the integration of research infrastructure and expertise, NMP-HEP has created a sustainable national framework for graduate education in a highly specialised scientific field.

The experience gained during the design and implementation of the programme highlights the value of institutional cooperation in addressing educational challenges that may be difficult to overcome by individual institutions acting alone. The establishment of NMP-HEP demonstrates that universities and research institutes can successfully combine complementary expertise and resources to create new educational opportunities while preserving their institutional autonomy. In this respect, the programme may provide a useful example for other disciplines and countries seeking to sustain advanced education in specialised fields under conditions of demographic decline and increasing international competition.

Although the long-term impact of the programme will only become evident after several generations of students have completed their studies, the successful completion of the first academic year provides encouraging evidence regarding the viability of this model. Looking ahead, NMP-HEP is expected to contribute not only to the preparation of a new generation of researchers capable of participating in major international scientific collaborations, but also to strengthening Romania's capacity for advanced research and graduate education in high-energy physics. The continued expansion of international partnerships, research opportunities, and doctoral pathways will further enhance the programme's contribution to both higher education and scientific research.

The present study provides a baseline for future evaluations of the programme. As the programme matures and increasing numbers of students complete their studies, it will become possible to assess its long-term impact through indicators such as graduation rates, national and international mobility, progression to doctoral studies, participation in international research collaborations, and employment outcomes.

Beyond its direct contribution to high-energy physics education, NMP-HEP demonstrates that national academic cooperation can provide an effective mechanism for preserving highly specialised disciplines whose long-term sustainability exceeds the capacity of individual institutions. The experience presented here suggests that coordinated national partnerships between universities and research institutes can offer a sustainable pathway for maintaining excellence in graduate education in highly specialised scientific disciplines.

*Acknowledgements*

The work of C.A., P.G., D.R., and R.Z. was supported by the Romanian Ministry of Education and Research through the ATLAS++ Contract. Z.I.L. acknowledges funding from the Romanian Ministry of Research, Innovation and Digitization, CNCS - UEFISCDI (project PN-IV-P2-2.1-TE-2023-1548) and the project ERANET-NEURON-2-RESIST-D.